\documentclass[sigconf,10pt]{acmart}
\newcommand{\nop}[1]{}
\newcommand{\rev}[1]{\textcolor{blue}{#1}} % 定义修改命令用于标注代码块
\newcommand{\sysname}{\textsf{GPUnion}}
\usepackage{xcolor}
\usepackage{booktabs}
\usepackage{threeparttable}

\title{\sysname{}: Autonomous GPU Sharing on Campus}

\author{Yufang Li}
\affiliation{
   \institution{HKUST(GZ)}%
   \country{}
 }

\author{Yuanbo Zhang}
\affiliation{
   \institution{Sun Yat-sen University}%
   \country{}
 }

\author{Hanlong Liao}
\affiliation{
   \institution{Sun Yat-sen University}%
   \country{}
 }

\author{Deke Guo}
\affiliation{
   \institution{Sun Yat-sen University}%
   \country{}
 }

\author{Guoming Tang}
\affiliation{
   \institution{HKUST(GZ)}%
   \country{}
 }

\begin{abstract}
A pronounced imbalance in GPU resources exists on campus, where some laboratories own underutilized servers while others lack the compute needed for AI research. GPU sharing can alleviate this disparity, while existing platforms typically rely on centralized oversight and persistent allocation models, conflicting with the voluntary and autonomous nature of academic resource ownership. 
We present \sysname{}, a campus-scale GPU sharing platform enabling voluntary participation while preserving full provider autonomy. 
\sysname{} incorporates three core mechanisms: i)~container-based task dispatching and execution, ii)~resource provider-first architecture, and iii)~resilient execution featuring automatic checkpointing and migration. 
\nop{\sysname{} also supports custom data storage and integrates the non-root execution and image attestation for isolation and security improvement for containerization.} 
\nop{Containerization offers lightweight deployment and workload portability, which are essential for voluntary participation; to compensate for its weaker isolation and security compared with virtual machines, we integrate non-root execution and image attestation, while retaining compatibility with lightweight VMs to keep the solution agile in trusted campus environments.}
Case studies across multiple campus scenarios demonstrate 30\% more GPU utilization improvement, 40\% increase in interactive sessions, and 94\% successful workload migration during provider departures. \nop{\sysname{} demonstrates that provider autonomy and platform reliability can coexist, challenging conventional centralized paradigms and democratizing access to computational resources within campus networks.}
\end{abstract}

\begin{document}

\acmYear{2025}\copyrightyear{2025}
\setcopyright{acmlicensed}
\acmConference[HotNets '25]{The 24th ACM Workshop on Hot Topics in Networks}{November 17--18, 2025}{College Park, MD, USA}
\acmBooktitle{The 24th ACM Workshop on Hot Topics in Networks (HotNets '25), November 17--18, 2025, College Park, MD, USA}
\acmDOI{10.1145/3772356.3772403}
\acmISBN{979-8-4007-2280-6/25/11}

\maketitle

\section{Introduction}

With the rapid development of artificial intelligence (AI), GPU resources have become a key infrastructure driving modern scientific computing and AI model serving. On communities like university campuses, however, GPU usage remains highly imbalanced. This imbalance manifests in multiple dimensions. 
(i)~Unequal resource distribution among faculties and departments, where some laboratories run sizeable GPU clusters while others have only minimal capacity. 
(ii)~Temporal underutilization, as research teams often experience significant idle periods between experiment cycles or during semester breaks. 
(iii)~Heterogeneous platform requirements, where different research groups need diverse GPU architectures, from consumer-grade RTX cards to high-end A100s, for varied computational workloads. 
(iv)~Accessibility barriers for students, particularly undergraduates and early stage researchers who lack institutional funding for dedicated hardware but require GPU resources for coursework and independent research projects. 

Ideally, a campus-wide GPU platform should democratize access to computing resources while improving overall utilization. Such a system would enable temporary borrowing of idle GPUs within a trusted network environment (e.g., campus LAN), reduce waste from hardware idleness, and support green research by maximizing the utility of existing devices instead of increasing hardware procurement. A platform like this would also foster collaboration and lower the cost of innovation for a broader academic community.

However, existing solutions fall short in supporting this vision. Comprehensive IaaS platforms like OpenStack~\cite{openstack_architecture} offer rich feature sets but impose prohibitive complexity and resource overhead for voluntary environments, requiring specialized expertise that conflicts with campus autonomy principles. While lightweight alternatives such as Apache CloudStack~\cite{kumar2014apache} and OpenNebula~\cite{milojivcic2011opennebula} reduce operational burden, their focus on traditional VM orchestration and rigid architectural assumptions fail to address the dynamic, containerized GPU sharing requirements of modern campus workloads. Container orchestration platforms like Kubernetes~\cite{kubernetes2019kubernetes}, despite supporting GPU scheduling through device plugins~\cite{senjab2023survey}, fundamentally rely on centralized control models that expect persistent node availability and stable connectivity, which is incompatible with the voluntary participation patterns in campus environments~\cite{kayal2020kubernetes}. Commercial cloud services such as AutoDL~\cite{autodl_platform} and academic cluster systems like Slurm~\cite{yoo2003slurm} operate on reservation based models that conflict with the spontaneous, revocable nature of campus resource sharing.

To this end, a dedicated GPU-sharing platform for campus environments is essential, and it must address several technical challenges. First, how to achieve \textit{secure isolation} while maintaining near-native GPU performance across heterogeneous hardware configurations without requiring extensive system administration. Second, how to design \textit{provider autonomy} mechanisms that allow resource owners to maintain full control over their machines while ensuring reliable service for users. Third, how to implement \textit{resilient execution} that transparently handles the inherent volatility of voluntary participation, including graceful migration and recovery from unexpected provider departures. Finally, how to create \textit{lightweight integration} that minimizes deployment friction.

In this paper, we propose \sysname{}, a campus-scale GPU sharing platform that enables voluntary resource contribution while preserving provider autonomy. \sysname{} is designed specifically for intra-campus environments, leveraging the trust and network proximity within local university networks. It employs containerized execution, strong host-guest isolation, and fault-tolerant mechanisms to support temporary, revocable, and secure GPU sharing. \nop{With \sysname{}, we demonstrate that campus-level GPU democratization is both technically feasible and practically beneficial.}

To be specific, we make the following major contributions:

\begin{itemize}
    \item \textbf{Containerized Execution Model}: We design a secure, lightweight execution framework based on OCI containers with GPU passthrough that provides near-native performance while maintaining strict isolation between guest workloads and host systems across heterogeneous hardware platforms.
    \item \textbf{Provider Supremacy Architecture}: We introduce a novel autonomy-first design that grants resource providers absolute control through kill-switch mechanisms and graceful departure protocols, enabling voluntary participation without sacrificing providers' experience.
    \item \textbf{Resilient Execution Mechanism}: We develop a comprehensive fault-tolerance system combining state-aware checkpointing and rapid migration capabilities to handle resource fluctuation in voluntary sharing environments. Users can specify specific nodes for data storage and backup according to their own needs.
\end{itemize}

\nop{Besides core GPU sharing capabilities, \sysname{} also supports advanced features such as cross-campus network federation, workload scheduling based on historical patterns, and integration with campus authentication systems. \nop{\sysname{} also provides a flexible storage mechanism that enables the configuration of user-defined target storage repositories. }We will open-source the entire \sysname{} platform under an academic-friendly license, providing detailed deployment guides and reference implementations for campus administrators and researchers.}

\section{Related Work}
The landscape of distributed computing platforms has evolved significantly to address various deployment scenarios and organizational needs. Understanding the strengths and limitations of existing solutions provides important context for positioning \sysname{}'s unique contribution\nop{to campus-scale resource sharing}.

\subsection{Infrastructure-as-a-Service Platforms}
OpenStack represents the most comprehensive approach to cloud infrastructure management, providing a full-featured IaaS platform with interrelated services for compute, networking, storage, and identity management~\cite{openstack_architecture}. While OpenStack's modular design and extensive ecosystem have proven successful at managing millions of cores in production clouds worldwide, its high complexity and heavy resource footprint make it unsuitable for the voluntary, lightweight sharing model required in campus environments. The platform requires significant expertise to deploy and operate, with steep learning curves for coordinating upgrades across multiple services, which conflicts with the autonomous participation principle central to \sysname{}.

\begin{table*}[!t]
\small
  \centering
  \setlength{\tabcolsep}{2pt}
  \begin{threeparttable}
    \caption{Comparison of Distributed Computing Platforms for Campus GPU Sharing}
    \label{tab:platform_comparison}
    \begin{tabular*}{\linewidth}{@{\extracolsep{\fill}} lccccc@{}}
      \toprule
      \textbf{Platform} &
      \textbf{OpenStack} &
      \textbf{CloudStack} &
      \textbf{OpenNebula} &
      \textbf{Kubernetes} &
      \textbf{GPUnion} \\
      \midrule
      \textbf{Community Support}           & Extensive & Limited & Limited & Extensive & Academic \\
      \textbf{Deployment Complexity}       & Very High & Medium  & Medium  & High      & Low \\
      \textbf{Resource Footprint}          & Very Heavy & Medium & Light   & Heavy     & Minimal \\
      \textbf{Learning Curve}              & Steep     & Moderate & Gentle  & Steep     & Gentle \\
      \textbf{Provider Autonomy}           & None      & None    & Limited & None      & Full \\
      \textbf{Workload Focus}              & VMs/Mixed & VMs     & VMs/Mixed & Containers & GPU Containers \\
      \textbf{Voluntary Participation}     & No        & No      & No      & No        & Yes \\
      \textbf{Dynamic Node Joining}        & Limited   & Limited & Limited & Limited   & Native \\
      \textbf{GPU Specialization}          & Add-on    & Limited & Add-on  & Plugin    & Core Feature \\
      \textbf{Campus Network Optimization} & No        & No      & No      & No        & Yes \\
      \textbf{Target Environment}          & Data Center & SME Clouds & Private Clouds & Large Clusters & Campus LANs \\
      \textbf{Fault Tolerance Model}       & Infrastructure & Infrastructure & Infrastructure & Infrastructure & Workload \\
      \bottomrule
    \end{tabular*}
  \end{threeparttable}
\end{table*}

Apache CloudStack takes a contrasting approach, emphasizing simplicity and quick deployment through a monolithic architecture where all core services come pre-integrated~\cite{kumar2014apache}. CloudStack's user-friendly design and support for multiple hypervisors make it more accessible than OpenStack for small-to-medium deployments. However, its less modular and extensible nature limits customization possibilities, while its smaller community and slower development pace restrict the availability of specialized features needed for dynamic GPU sharing scenarios.

OpenNebula positions itself as a lightweight alternative to OpenStack, designed for simplicity while maintaining enterprise-grade capabilities~\cite{milojivcic2011opennebula}. The platform supports both virtual machines and containers, offers hybrid cloud integration, and can be managed by small teams. Despite these advantages, OpenNebula's focus on traditional VM orchestration and its smaller developer community limit its applicability to the containerized, GPU-centric sharing model that \sysname{} requires.

\subsection{Container Orchestration Platforms}
Kubernetes has emerged as the dominant container orchestration platform, with approximately 96\% of organizations using or evaluating it for cloud-native workloads by 2022~\cite{Hardikar2021ContainerizationCCkubernetes_adoption}. Kubernetes provides powerful automation for container deployment, scaling, and recovery, along with extensive APIs and a rich ecosystem of add-ons. The platform supports GPU scheduling through device plugins and can accommodate diverse workloads through its flexible resource model.

However, Kubernetes' centralized control model fundamentally conflicts with principle that drives voluntary participation in campus environments. The platform expects stable connectivity and persistent node availability, poorly tolerating the dynamic participation patterns characteristic of voluntary resource sharing~\cite{kayal2020kubernetes}. Additionally, Kubernetes' complexity introduces operational overhead that exceeds the lightweight integration requirements of campus environments, where resource providers may have limited systems administration expertise.

\textbf{Positioning GPUnion.}
Table~\ref{tab:platform_comparison} summarizes the key characteristics of existing distributed computing platforms in comparison to \sysname{}'s requirements for campus-scale voluntary GPU sharing. Unlike existing platforms that prioritize either centralized control (OpenStack, Kubernetes) or simplicity through rigid architectures (CloudStack), \sysname{} introduces a novel governance model that preserves provider autonomy while ensuring system reliability. Our containerized execution approach combines Kubernetes' workload isolation benefits with a decentralized participation model that respects the voluntary nature of campus resource sharing. The platform's resilient execution mechanism specifically addresses GPU workload characteristics—checkpoint frequency optimization for intensive memory training, rapid migration for interactive sessions, and state-aware recovery for long-running jobs—capabilities absent from general purpose cloud platforms.

\sysname{} shares philosophical roots with earlier volunteer computing projects such as SETI@home~\cite{anderson2002seti} and Folding@home~\cite{pande2003atomistic}, where individuals contribute idle compute resources to large-scale scientific efforts. However, these systems typically operate in untrusted wide-area networks and rely on coarse-grained task splitting and result validation. In contrast, \sysname{} operates within a trusted campus LAN, enabling fine-grained, stateful workload migration and stronger accountability.
More importantly, while past volunteer systems focused on user-side incentives (e.g., leaderboard rankings, badges), \sysname{} shifts focus to provider-side empowerment. Instead of asking users to donate cycles, we ask labs to share underutilized infrastructure, making provider control not just a feature, but a foundational design principle. This reflects a broader trend toward community-driven resource pooling in academia~\cite{zaharia2010delay}.

\section{GPUnion Design}

Building a cooperative GPU sharing platform within a campus network requires reconciling a set of conflicting goals: i)~giving resource providers full autonomy while ensuring reliability for resource users, ii)~maintaining security while avoiding administrative burden, and iii)~achieving broad compatibility while minimizing system overhead. To guide the design of \sysname{}, we set the following core principles.

\subsection{Design Principles}
\begin{itemize}
\item \textbf{Autonomous Participation.}  
Unlike data center clusters, campus resources are inherently decentralized and independently managed. \sysname{} adopts an autonomy first paradigm. Every server node can independently decide when to join, pause, or leave the network. This ensures that faculty or lab-owned servers, personal workstations, or idle lab GPUs can be shared on a voluntary and revocable basis. Each resource provider retains full authority, and the platform operates in a non-intrusive manner, respecting individual ownership boundaries.

\item \textbf{Lightweight and Transparent Usage.}  
To encourage adoption among both contributors and consumers, \sysname{} is designed to be frictionless. For users, submitting a job to the system should feel no more complex than running it locally, while the underlying scheduling, migration, and checkpointing remain invisible. For providers, hosting workloads should not require reconfiguring their machines or permanently dedicating resources. The platform is based on containerization technology, enabling seamless encapsulation of workloads with minimal performance overhead and strong host-guest isolation.

\item \textbf{Reliability, Resilience, and Security.}  
Given the voluntary and dynamic nature of participation, the system must be robust against unpredictable resource churn. \sysname{} is designed to ensure user workloads remain resilient to interruptions through automatic checkpointing and migration mechanisms. At the same time, it enforces strict sandboxing for all external workloads, ensuring that guest code cannot access or interfere with the host system. These mechanisms uphold user trust and maintain system stability despite fluctuating resource availability.
\end{itemize}

\subsection{GPUnion Overview}
\begin{figure}[t]
\centering
\includegraphics[width=0.48\textwidth]{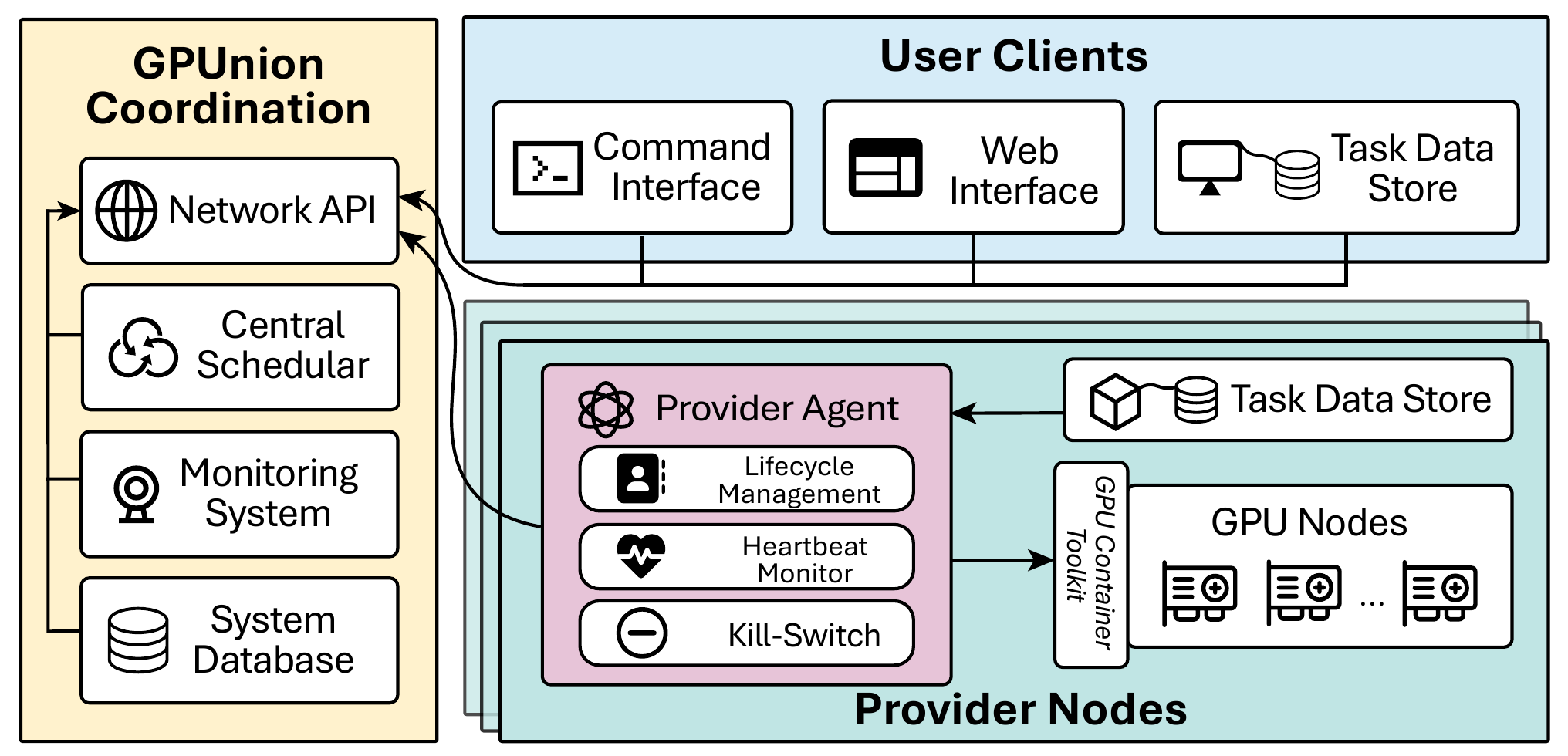}
% \vspace{-0.5cm}
\caption{GPUnion system architecture diagram.}
\label{fig:system_architecture}
\vspace{-0.5cm}
\end{figure}

\nop{Based on the above design principles, \sysname{} is composed of three key components that work together in coordination. }This section provides an overview of \sysname{} system architecture, which is also illustrated by Fig.~\ref{fig:system_architecture}.

\nop{The design of \sysname{} centers on creating a resilient, cooperative GPU-sharing platform within a campus LAN, where resource providers retain full autonomy and trust is prioritized over centralized enforcement. The system architecture comprises three key components that work together to enable voluntary GPU sharing while preserving provider control and ensuring reliable service delivery.
}

\textbf{Central Scheduler and Coordinator.} The central scheduler serves as the coordination hub for resource discovery, allocation decisions, and workload management. It maintains a real-time view of available GPU resources across the campus network through periodic status updates from provider agents. The scheduler implements multiple allocation strategies, including distribution for fairness and assignment based on priority for time-sensitive workloads. Unlike traditional cluster schedulers that assume persistent resource availability, \sysname{}'s scheduler is designed to handle dynamic resource volatility, incorporating provider reliability predictions and \rev{\nop{graceful}} degradation mechanisms.

\textbf{Provider Agents and Local Control.} Each participating node runs a lightweight agent that implements the provider supremacy model through local control mechanisms and real-time monitoring. The agent exposes REST APIs for resource advertisement, workload lifecycle management, and emergency controls while maintaining absolute provider authority through ``kill-switch'' functionality. Providers can instantly terminate running workloads, pause further allocations, or disconnect from the platform without coordination overhead. The agent automatically handles node registration, authentication token management, and secure network connectivity through the network API.

\textbf{Containerized Execution Environment.} All GPU workloads execute within Docker containers configured with NVIDIA Container Toolkit for direct GPU access. \nop{This containerization layer provides workload portability across heterogeneous campus hardware while maintaining strict isolation through Linux kernel primitives including namespaces, cgroups, and Seccomp profiles.} The system supports both interactive research environments \rev{\nop{(automatically provisioned Jupyter notebooks with pre-configured deep learning frameworks)}}and batch execution modes for production workloads with customizable container images and entry points.

\textbf{Distributed State Management and Monitoring.} A comprehensive monitoring and state management system collects real-time telemetry from all participating nodes through metrics exporters. The system captures both hardware metrics and application metrics. State persistence is handled through a centralized\rev{\nop{PostgreSQL}} database that maintains node registrations, resource allocations, and historical monitoring data, enabling both operational decision making and capacity planning.

\textbf{Data Storage Architecture.} \sysname{} implements a flexible data storage model that accommodates both centralized and distributed storage needs across client and provider environments. On the client side, users can specify preferred storage locations for their workload data, checkpoints, and outputs, enabling them to maintain control over sensitive datasets while using distributed compute resources. Provider servers offer local storage capabilities for temporary data and intermediate results, while supporting integration with campus-wide distributed file systems for persistent storage.

\textbf{Technical Challenges.} Designing \sysname{} introduces three core technical challenges. First, achieving secure and performant execution across heterogeneous hardware requires a containerization solution that delivers near-native GPU performance while maintaining strong isolation, despite variations in drivers, OS configurations, and security policies. Second, preserving provider autonomy, such as through an immediate kill-switch, must be reconciled with user experience, as abrupt resource reclamation risks disrupting ongoing workloads, necessitating graceful termination protocols that balance control and fairness. Third, ensuring resilient execution under arbitrary node departures demands transparent checkpointing and fast migration mechanisms, since provider-initiated exits make failure prediction impossible and workload continuity must be maintained without relying on infrastructure-level redundancy.

\nop{In summary, \sysname{} introduces three core mechanisms: containerized execution for isolation, provider supremacy for dynamic autonomy, and a resilient execution mechanism that mitigates volatility in participation. Together, these mechanisms turn a collection of heterogeneous and individually managed machines into a logically unified compute platform.} \nop{that respects individual ownership while maximizing collective utility.} 
\nop{Fig.~\ref{fig:system_architecture} presents the overall system architecture and component interactions. When discussing resource management, we focus on single-node participation; more complex scenarios involving heterogeneous or centrally-managed clusters are left for future work.}

\subsection{Containerized GPU Execution}
To achieve transparent and lightweight usage across a wide variety of machines, \sysname{} adopts containerized execution based on OCI standards (e.g., Docker) combined with GPU passthrough using the NVIDIA Container Toolkit. Each job is deployed inside an isolated user-space container, leveraging Linux kernel primitives such as namespaces, cgroups, and Seccomp profiles to ensure strict resource boundaries.

This approach offers several advantages. First, it provides near-native GPU performance by allowing user workloads to access the GPU directly, avoiding the overhead of full virtualization. Second, it guarantees security and isolation, preventing any container from affecting the host environment or neighboring workloads. Third, it ensures portability: Containers can be deployed on a wide range of machines regardless of driver or OS variations, a critical feature given the hardware diversity in campus networks.

\textbf{Implementation Details.} Our containerization layer supports both interactive and batch execution modes through standardized Docker runtime environments configured with GPU passthrough capabilities. For interactive research, the system automatically provisions Jupyter notebook environments with pre-configured deep learning frameworks and GPU access through the NVIDIA Visible Devices environment variable. For production workloads, it supports arbitrary container images with customizable entry points and environment configurations. Container images must pass SHA256 verification before deployment, and the system maintains an allow list of trusted base images to ensure security compliance.

\subsection{Provider Supremacy and Autonomy}
A core principle of \sysname{} is the rejection of centralized control in favor of provider supremacy. In contrast to traditional cluster or edge platforms, where compute nodes are centrally managed and statically allocated, \sysname{} enables each provider to freely join, leave, or pause participation at any time, without prior negotiation.

This autonomy is enabled by a lightweight agent running on each provider's machine. The agent exposes the machine's availability status and GPU capacity to the central scheduler, but always allows the provider to immediately override the system via a local "kill-switch." At any point, a provider can terminate running workloads, pause further task scheduling, or disconnect entirely.

\textbf{Implementation Details.} 
Each provider node runs a lightweight agent that implements provider supremacy by building an API server to handle resource advertisement, workload lifecycle management, and emergency controls. The agent integrates with PyNVML to collect real-time GPU telemetry including memory utilization, temperature, and power consumption. Provider supremacy is implemented through ``kill-switch'' mechanism. When providers trigger voluntary exit, the agent performs workload termination with configurable\rev{\nop{grace}} periods for checkpoint creation. New nodes join the platform through automatic registration scripts that generate unique machine identifiers, establish network connectivity, and obtain authentication credentials.

\subsection{Resilient Execution Mechanism}
Because provider autonomy introduces inherent volatility, \sysname{} must ensure task continuity in a setting where nodes may leave arbitrarily. To address this, the platform incorporates a resilient execution mechanism with the following capabilities:

\textbf{State-Aware Checkpointing.} For long-running or stateful workloads, the platform periodically captures full application state—CPU and memory snapshot, file system state, and GPU memory if feasible—via container-level checkpointing backup. These checkpoints can be stored in a LAN-accessible \rev{\nop{distributed}}file system or a specific node.

\textbf{Rapid Migration and Recovery.} When a provider voluntarily exits or fails, the central scheduler is notified via heartbeat loss. The workload is then automatically relaunched on another available machine using the latest checkpoint. For stateless tasks, the process simply involves requeuing and redispatching the job. In both cases, \sysname{} hides provider side volatility from end-users, offering a consistent service experience.

\textbf{Implementation Details.} The resilient execution mechanism operates through a modular architecture supporting multiple allocation strategies via a round-robin scheduler (which processes pending resource requests from a priority queue stored in the central database). Resource allocation decisions consider GPU memory requirements, CUDA compute capability constraints and provider volatility predictions\rev{\nop{, and network latency metrics}}. The system implements heartbeat-based failure detection with configurable timeouts, i.e., nodes that miss three consecutive heartbeats are marked as unavailable, triggering automatic workload migration. Comprehensive monitoring is achieved through Prometheus metrics exporters that collect both hardware metrics (GPU utilization, memory usage, temperature, etc.) and application metrics (container lifecycle events, resource allocation history, etc.) for real-time operational decisions and historical capacity planning.

\sysname{} ensures continuity during provider departure via application-level checkpoints (ALC). On voluntary exit, workloads save user-specified state to designated storage. The scheduler then relaunches the task on a compatible node with the latest checkpoint, enabling fast and reliable migration with minimal downtime.

A critical challenge in building a resilient yet lightweight sharing platform lies in choosing the right mechanism for state preservation and migration. We evaluated several alternatives before settling on ALC as the cornerstone of our design. System level solutions like CRIU (Checkpoint/Restore in Userspace), while powerful, fail to support CUDA contexts reliably and impose strict requirements on kernel versions and driver compatibility. More importantly, they cannot support cross-GPU architecture migration, which is common in heterogeneous campus environments. Alternative replication schemes based on P2P would require complex coordination logic and additional networking layers, contradicting \sysname{}'s goal of minimal intrusion. In contrast, ALC shifts responsibility to users who already manage their own training scripts and data persistence to define what constitutes recoverable state. This aligns with real world AI development practices where models are routinely saved incrementally, and thus reduces system complexity while increasing practicality across diverse hardware configurations.

%% Case study content on the performance of GPUnion in a real-world
%% campus testbed will be added here. We plan to evaluate workload
%% migration latency, checkpointing overhead, and the platform's
%% ability to support large-scale distributed training tasks.

\section{Case Studies}

We deploy \sysname{} in a campus network environment comprising 11 GPU services. Among these, 8 servers functioned as workstations, each equipped with a single NVIDIA 3090 GPU; one server featured 8 4090 GPUs; another two servers housed 2 A100 and 4 A6000, respectively. An additional CPU-only server served as the central coordinator. Prior to the deployment, all resources are managed through manual coordination. Each server is equipped with the \sysname{} agent alongside existing job management tools, allowing for a smooth transition and incremental adoption. \nop{Resource providers retained full control through local kill-switch mechanisms, while the central scheduler operated as an auxiliary resource discovery module.} 

\begin{figure}[t]
\centering
\includegraphics[width=0.45\textwidth]{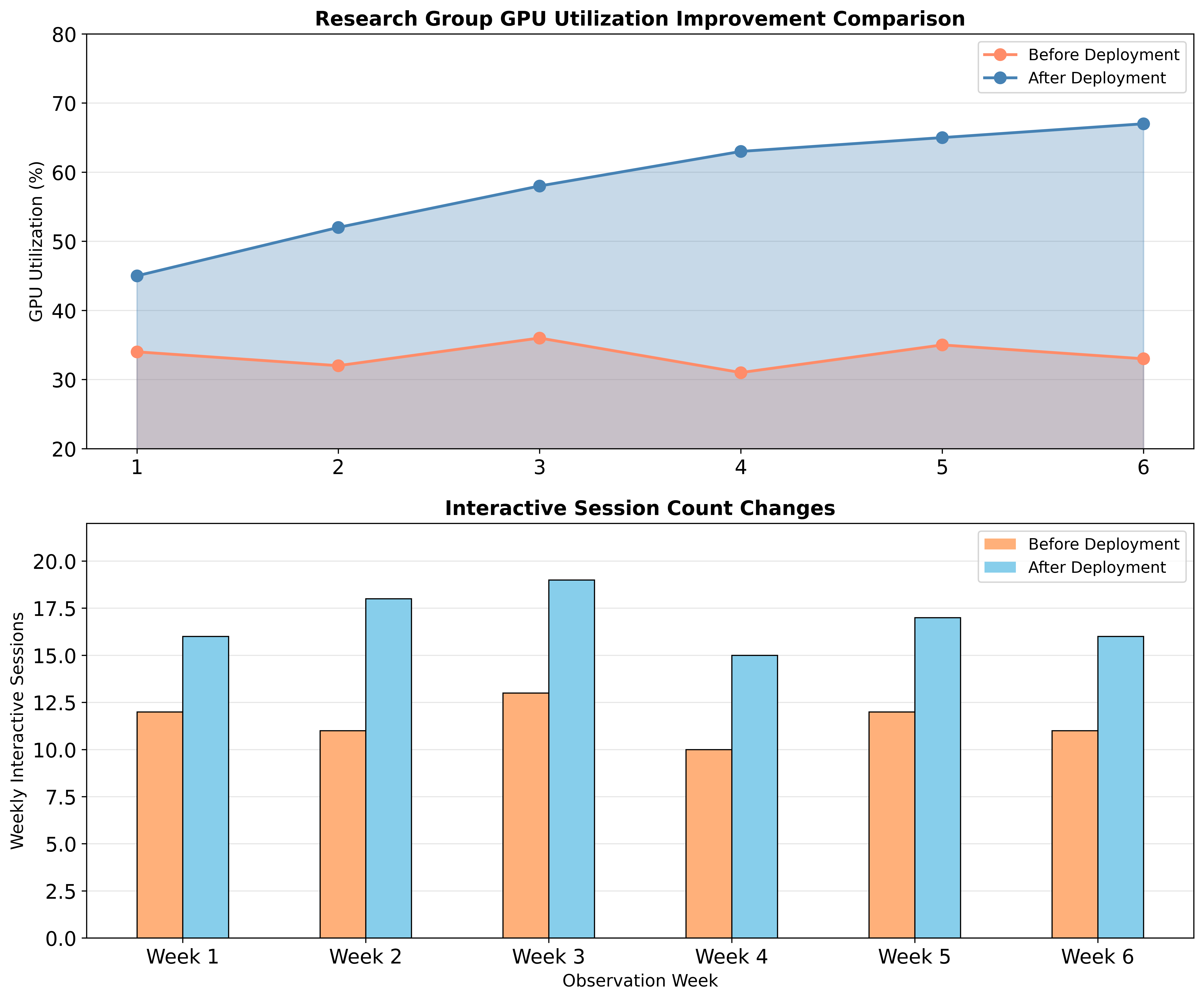}
% \vspace{-0.45cm}
\caption{Research group GPU utilization comparison.}
\label{fig:gpu_utilization}
\vspace{-0.3cm}
\end{figure}

As shown in Fig.~\ref{fig:gpu_utilization}, after a six-week period, the average GPU utilization of all servers increased from 34\% to 67\%. This improvement was primarily attributed to enhanced visibility of resource availability and the automated allocation of opportunistic workloads during idle periods. Furthermore, interactive debugging sessions increased by 40\% compared to the manual coordination phase, as students were able to access temporarily idle GPUs more conveniently.

To evaluate \sysname{}'s resilience mechanisms, we conducted controlled experiments simulating realistic provider interruption patterns. These experiments involved 20 deep learning training jobs (PyTorch CNN and transformer models) distributed across 2 volunteer provider nodes over a week period.

\begin{figure}[t]
\centering
\includegraphics[width=0.45\textwidth]{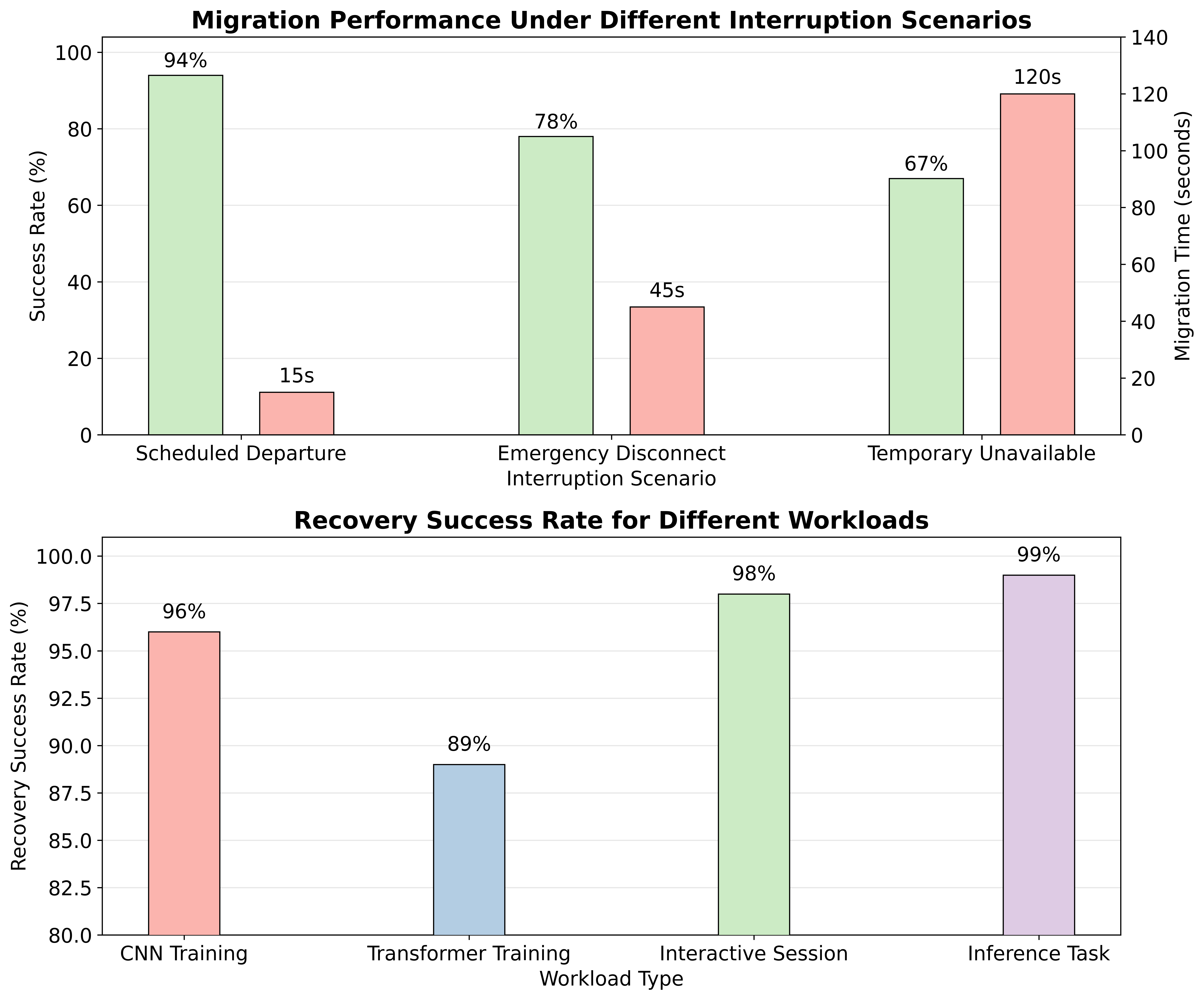}
\caption{Migration performance under different interruption scenarios.}
\label{fig:migration_performance}
\vspace{-0.5cm}
\end{figure}

\textbf{Interruption Scenarios:} We simulated three classes of provider behavior: \textit{scheduled departure} (provider initiates graceful shutdown), \textit{emergency departure} (immediate disconnection), and \textit{temporary unavailability}. Interruption frequency varied from 0.5 to 3.2 events per day per node, reflecting realistic campus usage patterns. For scheduled departures, 94\% of workloads successfully migrated within the specified time and with minimal data loss. Emergency departures resulted in work loss equivalent to the checkpoint interval. Temporary unavailability scenarios demonstrated the value of provider return 67\% of displaced workloads were automatically migrated back to their original nodes in time when providers reconnected. Fig.~\ref{fig:migration_performance} provides a comprehensive analysis of these migration performance characteristics across different interruption scenarios and workload types.

\textbf{Training Impact:} Despite frequent interruptions, training convergence was minimally affected. Jobs experiencing 2-4 interruptions showed only 3-7\% increases in total training time compared to uninterrupted execution. Memory-intensive models showed higher sensitivity to interruption due to longer checkpoint creation times, suggesting the value of workload-specific checkpoint strategies.

\textbf{Network Traffic Analysis:} Given that frequent checkpointing and migration operations could potentially saturate campus network infrastructure and interfere with other critical applications, we conducted comprehensive network traffic analysis during backup operations. Our measurements across various workload types revealed that the incremental checkpointing mechanism produces negligible network overhead, with backup traffic consuming less than 2\% of available campus bandwidth during peak operation periods. The incremental nature of state synchronization—where only modified memory pages and file system deltas are transmitted—ensures that \sysname{}'s resilience mechanisms operate transparently without impacting concurrent network-intensive research activities or administrative services.

\section{Discussions and Opportunities}

\subsection{Rethinking the Novelty of \sysname{}}

While individual components of GPUnion (containerization, checkpointing, and agent-based coordination) are built upon existing technologies, its novelty lies in the recombination around a new design principle: provider autonomy as the foundation of resource sharing. The "kill-switch" is not merely a feature but an embodiment of this principle, enabling providers to instantly reclaim resources without negotiation or penalty. This contrasts sharply with traditional platforms (e.g., Kubernetes, Slurm), where nodes are expected to remain stable and available, often under centralized administrative control. In those systems, volatility is treated as failure; in GPUnion, it is first-class behavior.

Why has such a mechanism not been widely adopted before? Prior efforts in cloud and cluster computing prioritize service reliability over provider flexibility, assuming either commercial incentives or institutional mandates for resource contribution. In contrast, campus environments operate on trust and mutual benefit rather than payment. Here, the marginal cost of electricity and maintenance is low, reducing economic barriers to sharing. Yet, psychological and operational barriers remain high: researchers hesitate to share if they cannot immediately regain control during urgent experiments. By placing autonomy at the center, GPUnion reframes voluntary sharing not as a technical problem of scheduling, but as a socio-technical alignment between trust, convenience, and utilization. We find that the simplest mechanism, the kill switch, is precisely what enables adoption, demonstrating that in trusted settings, minimalism can be more powerful than complexity.

\subsection{Opportunities in the Future}

\sysname{} also opens several research avenues that address fundamental challenges in distributed computing and voluntary resource sharing. \nop{These directions span from immediate deployment considerations to long-term architectural innovations.}

\textbf{Heterogeneous Large Model Deployment.} \nop{GPUnion's heterogeneous resource environment presents unique opportunities for advancing large model deployment strategies. Unlike homogeneous clusters, campus networks contain diverse GPU architectures with varying memory capacities, compute capabilities, and interconnect bandwidths. This heterogeneity requires novel approaches to model partitioning, layer assignment, and load balancing that consider both hardware constraints and provider volatility patterns. Future research should explore adaptive partitioning algorithms that dynamically reassign model segments based on real-time resource availability, develop memory-aware sharding strategies, and investigate mixed-precision training techniques.} 
The heterogeneous resource environment in which \sysname{} operates offers unique opportunities for large model deployment. Unlike homogeneous clusters, \sysname{} deploys in campus networks, which host a variety of GPU architectures whose memory capacity, compute capability, and interconnect bandwidth differ substantially. This heterogeneity calls for new approaches to model partitioning, layer placement, and load balancing that simultaneously respect hardware constraints and the fluctuating availability of contributors.

\rev{\nop{\textbf{Incentive Mechanism Design for GPU Node Contribution}  
An effective incentive mechanism is key to the sustainable operation of \sysname{}. Future work can proceed on multiple fronts. For example, establishing a compute-time–based credit system and designing dynamic auction or floating-rate schemes that reflect actual supply and demand for different GPU architectures, memory sizes, and network bandwidths. Given the non-profit nature of the campus, our aim is to encourage more GPU contributions while ensuring that resource-seeking users are not deterred by an overly complex framework.}}

\textbf{User-Transparent Resource Invocation.} \sysname{} currently requires users to estimate their own resource needs and then request those resources. This process is cumbersome, and inaccurate estimates can easily lead to resource waste. Exposing \sysname{} through a programming interface, such as a Python package, and incorporating intelligent mechanisms for resource estimation, requesting, and scheduling would greatly improve both efficiency and utilization.

\textbf{Scalability.} \sysname{} is designed for mid-sized campus environments (tens to hundreds of GPUs), where trust and proximity enable lightweight coordination. In our deployment, the central coordinator handles up to 50 nodes with sub-second scheduling latency. However, beyond ~200 nodes, heartbeat monitoring and database contention could become bottlenecks. For larger deployments, future work will explore hierarchical coordination or gossip-based decentralization while preserving autonomy.
We believe there exists a "sweet spot": large enough to benefit from resource pooling, yet small enough to maintain low-latency communication and social cohesion among providers.

\section{Conclusion} 

This paper presented \sysname{}, a campus-scale GPU-sharing platform to prioritize the autonomy of resource providers in a volunteer computing environment. The main features of \sysname{} including a container-based execution model that strikes a balance between performance and security in heterogeneous hardware settings, a provider-centric design that allows providers to reclaim their GPUs instantly through a kill-switch mechanism, and an elastic execution mechanism that offers transparent fault tolerance.
We validated the platform in a campus scenario, and the results show significant benefits in real deployments.\nop{When provider autonomy is properly safeguarded, the volunteer-participation model can likewise achieve both high performance and service reliability.}
\sysname{} demonstrates that provider autonomy and platform reliability can coexist, challenging conventional centralized paradigms and democratizing access to scattered computational resources.

\section*{Acknowledgments}

We thank the reviewers and our shepherd for their valuable and insightful feedback. We are also grateful to our lab members, Hemu Liu, Kunming Zhang, Junyu Xue, and Yuxi Zhao, for their support and fruitful discussions during the development of \sysname{}. This work was supported in part by the National Natural Science Foundation of China under Grant No. U23B2004.

\bibliographystyle{ACM-Reference-Format} 
\bibliography{refs}

\end{document}